# Feinberg-Horodecki States of Time-Dependent Mass Distribution Harmonic Oscillator

M. Eshghi [1,*], R. Sever [2], S. M. Ikhdair [3,4]

[1] *Researchers and Elite Club, Central Tehran Branch, Islamic Azad University, Tehran, Iran*

[2] *Physics Department, Middle East Technical University, 06530 Ankara, Turkey*

[3] *Department of Physics, Faculty of Science, An-Najah National University, Nablus, West Bank Palestine*

[4] *Department of Electrical Engineering, Near East University, Nicosia, Northern Cyprus, Mersin 10, Turkey*

**Abstract**

The solution of the Feinberg-Horodecki (FH) equation for a time-dependent mass (TDM) harmonic oscillator quantum system is studied. A certain interaction is applied to a mass $m(t)$ to provide a particular spectrum of stationary energies. The related spectrum of the harmonic oscillator potential $V(t)$ acting on the TDM $m(t)$ oscillators is found. We apply the time version of the asymptotic iteration method (AIM) to calculate analytical expressions of the TDM stationary state energies and their wave functions. It is shown that the obtained solutions reduce to those of simple harmonic oscillator as the time-dependent of the mass reduces to $m_0$.

---

[*] Corresponding author mail: eshgi54@gmail.com; m.eshghi@semnan.ac.ir

## 1. Introduction

It is interesting to find the exact analytical bound state solutions of the Feinberg-Horodecki (FH) equation for the time-dependent mass under the harmonic oscillator potential interaction.

On the other hand, systems with time-dependent mass (TDM) have been found to be very useful in studying different fields. In this regards, different potential models are widely used in many branches of physics (see for example [1,2,3,4]). In the quantization of fields, in curved space-times [5], in cosmology [6, 7], and the description of the scalar fields [8].

On the other hand, the time-dependent mass has been studied in different systems. Here, the time-dependent mass distribution of any oscillating systems has its importance in the ontogenic growth [9] in the so-called the West, Brown and Enquist (WBE) function. This function has the form: $y(t) = [1 - c_3 \exp(-c_1 t)]^{1/c_2}$, where $y(t) = m(t)/M$, $c_3 = 1 - (m_0/M)^{1/c_2}$, $c_1 = n_1/4M^{c_2}$, $c_2 = 1 - 3/4$, $n_1 = B_0 m_c/E_c$, $B_0$ is the initial ($t=0$) average resting metabolic rate of the whole organism, $E_c$ is the metabolic energy required to form a cell, $m_c$ is the mass of a cell, $m_0 = m(t=0)$ is the initial mass of the system whereas $M = m(t \to \infty) = (n_1/n_2)^{1/c_2}$ is the maximum body size reached where $n_2 = B_c/E_c$ with $B_c$ is the metabolic rate of a single cell and $dm(t)/dt = n_1 m^{3/4} - n_2 m$. Also, in this regard, Dantas et al have investigated the harmonic oscillator with time-dependent mass along with frequency and a perturbative potential [10], Ji and Kim have found the exact quantum motions in the Heisenberg picture for a harmonic oscillator with time-dependent mass and frequency in terms of classical solutions [11], Maamache *et al* have generalized unitary equivalence and phase properties of the quantum parametric in harmonic oscillators [12], Axel Schulze-Halberg has studied the quantum systems with effective and time-dependent masses [13], Lai et al have investigated the many-body wave function of a quantum system with time-dependent effective mass, confined by a harmonic potential with time-dependent frequency, and perturbed by a time-dependent spatially homogeneous electric field [14]. Moya-Cessa and Fernandez Guasti have also studied the time dependent quantum harmonic oscillator that subjects to a sudden change of mass: continuous solution [15] and so forth.

The time-like counterpart of the Schrödinger equation was derived by Horodecki [16] from the relativistic Feinberg equation [17]. However, in the case of harmonic vector potential with time-dependent mass there are no available bound states in the dissociation limit and the direction of temporal motion is consistent with the arrow of time. In such circumstances, the time-like solutions of the FH equation is employed to test their relevance in physics, biology and medicine [18]. Using the Nikoforov-Uvarov method, some of the authors have investigated the FH equation with the time-dependent different potentials [1, 19]. Also, Molski has investigated this equation for interpreting the formation of the specific growth patterns during crystallization process and biological growth [18] and the biological systems with anharmonic oscillators [9]. On the other hand, in recent years, many authors have investigated coherent states for different potentials [20-23]. These cases are made using a transformation of the basic Hamiltonian into a resembling form by an algebraic method [24].

In what follows, we first construct the space-like coherent states with the time-dependent harmonic oscillator potential and then the mass time distribution of the FH equation.

Thus, using some well-known appropriate methods, we obtain the energy eigenvalues and the corresponding wave functions by using the time version of Asymptotic Iteration Method (AIM).

The paper is organized as follows. In Sec. 2 we briefly introduce the time version of AIM. In section 3, we obtain the bound energy eigenvalues and the corresponding wave functions of the FH equation with time-dependent mass time harmonic oscillator potential by using the the time version of the AIM. Finally we give our discussions and concluding remarks in Sec.4.

## 2   Time Like Asymptotic Iteration Method (AIM)

Our aim, in the present work, is the extension of the space-like AIM in order to include time-like AIM, which we shall use in obtaining solutions of the FH equation. To achieve this goal, we substitute $x$ parameter with $t$ parameter in whole following equations. In other words, in the time-like version of AIM, we assume the space parameter, $x$, to be as time parameter, $t$. Therefore, we briefly introduce the time-like AIM here; the detail of the method can be traced from Refs. [25, 26]

$$y_n''(t) = \lambda_0(t) y_n'(t) + s_0(t) y_n(t), \quad \lambda_0(t) \neq 0, \tag{1}$$

where $\lambda_0(t)$ and $s_0(t)$ are sufficiently differentiable parameters. To find a general solution to this equation, we start differentiating Eq. (1) as

$$y_n'''(t) = \lambda_1(t) y_n'(t) + s_1(t) y_n(t), \tag{2}$$

where the assignments

$$\lambda_1(t) = \lambda_0'(t) + s_0(t) + \lambda_0^2(t), \tag{3a}$$

and

$$s_1(t) = s_0'(t) + s_0(t)\lambda_0(t), \tag{3b}$$

have been used. Further, we proceed by taking the second derivative of Eq. (1) as

$$y_n^{(4)}(t) = \lambda_2(t) y_n'(t) + s_1(t) + \lambda_0(t)\lambda_1(t), \tag{4}$$

where

$$y_2(t) = \lambda_1'(t) + s_1(t) + \lambda_0(t)\lambda_1(t), \tag{5a}$$

and

$$s_2(t) = s_1'(t) + s_0(t)\lambda_1(t). \tag{5b}$$

We repeat to $(k+1)th$ derivatives with $k = 1,2,3,...$ and obtain

$$y_n^{(k+1)}(t) = \lambda_{k-1}(t) y_n'(t) + s_{k-1}(t) + y_n(t), \tag{6}$$

and

$$y_n^{(k+2)}(t) = \lambda_k(t) y_n'(t) + s_k(t) y_n(t),$$

where

$$s_k(t) = s_{k-1}'(t) + s_0(t)\lambda_{k-1}(t). \tag{7}$$

The ratio of the $(k+2)th$ and $(k+1)th$ derivatives gives

$$\frac{d}{dt}\ln\left[y_n^{k+1}(t)\right] = \frac{y_n^{(k+2)}(t)}{y_n^{(k+1)}(t)} = \frac{\lambda_k(t)\left[y_n'(t) + \frac{s_k(t)}{\lambda_k(x)} y_n(t)\right]}{\lambda_{k-1}(t)\left[y_n'(t) + \frac{s_{k-1}(t)}{\lambda_{k-1}(t)} y_n(t)\right]}, \qquad (8)$$

and hence for sufficiently large $k$ [20,21]

$$\frac{s_k(t)}{\lambda_k(t)} = \frac{s_{k-1}(t)}{\lambda_{k-1}(t)} = \alpha(t), \qquad k = 1,2,3,.... \qquad (9)$$

with the quantization condition

$$\begin{vmatrix} \lambda_k(t) & s_k(t) \\ \lambda_{k-1}(t) & s_{k-1}(t) \end{vmatrix} = 0, \qquad k = 1,2,3,.... \qquad (10)$$

Thus, the general solution of Eq. (1) takes the form

$$y_n(t) = e^{-\int^t \alpha(t')dt'}\left[C_2 + C_1 \int^t e^{\left(\int^{t'}[\lambda_0(t'')+2\alpha(t'')]dt''\right)} dt'\right], \qquad (11)$$

where $C_1$ and $C_2$ are two constants.

## 3 The Feinberg-Horodecki Equation

The space-like counterpart of the Schrödinger equation called the FH equation which can be casted in the form:

$$-\frac{\hbar^2}{2}\frac{d}{dt}\left[\frac{1}{m(t)}\frac{d}{dt}\varphi(t)\right] + V(t)\varphi(t) = E_n\varphi(t), \quad t = x_0/c. \qquad (12)$$

where the interaction potential model is taken as the time-dependent harmonic oscillator potential:

$$V(t) = \frac{1}{2}m(t)\omega^2 t^2, \qquad (13)$$

where $\omega$ is its natural frequency. Here, we use the following particular choice for the mass time distribution as

$$m(t) = \frac{m_0}{1 + \lambda t^2}, \qquad (14)$$

where $\hbar = c = m_0 = 1$, leads to the celebrated nonlinear oscillator [27, 28].

Now, with respect to space and time coordinates, the space- and time-like coherent states and equations governing their propagation are symmetric. In fact, in the case of stationary states with the momentum $P=const$, the space-independent version of the FH equation is drivable by proper substitution as $\varphi = \phi(t)\exp(-iPx/\hbar)$ into Eq. (12), we have

$$-\frac{\hbar^2}{2}\frac{d}{dt}\left[\frac{1}{m(t)}\frac{d}{dt}\phi(t)\right] + V(t)\phi(t) = P\phi(t).$$

This is identical to an energy eigenvalue equation in a scalar potential, with $x$ replaced by $t$ [16].

When Eqs. (13) and (14) are substituted ont0 Eq. (12), we have

$$-\frac{1}{2}(1+\lambda t^2)\frac{d^2\varphi(t)}{dt^2} - \lambda t\frac{d\varphi(t)}{dt} + \frac{1}{2}\left(\frac{\omega^2 t^2}{1+\lambda t^2}\right)\varphi(t) = E_n\varphi(t). \quad (15)$$

After simplifying Eq. (15), we can introduce the following dimensionless quantities:

$$\tau = \sqrt{\omega}t, \quad \tilde{\lambda} = \lambda/\omega, \quad \tilde{E} = 2E/\omega, \quad (16)$$

to obtain

$$(1+\tilde{\lambda}\tau^2)\frac{d^2\varphi(\tau)}{d\tau^2} + 2\tilde{\lambda}\tau\frac{d\varphi(\tau)}{d\tau} + \left(\tilde{E} - \frac{\tau^2}{1+\lambda\tau^2}\right)\varphi(\tau) = 0. \quad (17)$$

Now, we choose an appropriate ansatz form for the wave function at short, long and intermediate ranges as

$$\varphi(\tau,\tilde{\lambda}) = (1+\tilde{\lambda}\tau^2)\exp\left(-\frac{1}{2}\tilde{\lambda}\right)f(\tau,\tilde{\lambda}), \quad (18)$$

so that Eq. (17) turns out to become

$$(1+\tilde{\lambda}\tau^2)\frac{d^2 f(\tau,\tilde{\lambda})}{d\tau^2} + 2(1-\tilde{\lambda})\tau\frac{df(\tau,\tilde{\lambda})}{d\tau} + (\tilde{E}-1)f(\tau,\tilde{\lambda}) = 0. \quad (19)$$

Now, our aim is to solve the above equation by means of time-like AIM given in Sec. 2. Further, when comparing Eq. (19) with Eq. (1), we can extract the values of the parameters $\lambda_0(\tau)$ and $s_0(\tau)$ as

$$\lambda_0 = -\frac{2(1-\tilde{\lambda})\tau}{1+\tilde{\lambda}\tau^2}, \quad s_0 = \frac{1-\tilde{E}}{1+\tilde{\lambda}\tau^2}, \quad (20)$$

and also seeking Eq. (9), we can obtain the parameters $\lambda_n(\tau)$ and $s_n(\tau)$ for $n=1$ as

$$\lambda_1 = \frac{2(1-\tilde{\lambda})}{1+\tilde{\lambda}\tau^2} - \frac{4(1-\tilde{\lambda})\tau^2\tilde{\lambda}}{(1+\tilde{\lambda}\tau^2)^2} + \frac{1-\tilde{E}}{1+\tilde{\lambda}\tau^2} + \frac{4(1-\tilde{\lambda})^2\tau^2}{(1+\tilde{\lambda}\tau^2)^2}, \tag{21a}$$

and

$$s_1 = -\frac{2(1-\tilde{E})\tilde{\lambda}\tau}{1+\tilde{\lambda}\tau^2} + \frac{2(1-\tilde{\lambda})(1-\tilde{E})\tau}{(1+\tilde{\lambda}\tau^2)^2}. \tag{21b}$$

Repeating this iteration routine, we can hence obtain these parameters for $n \geq 2$ as

$$\lambda_2 = -\frac{12(1-\tilde{\lambda})\tilde{\lambda}\tau}{(1+\tilde{\lambda}\tau^2)^2} + \frac{16(1-\tilde{\lambda})\tilde{\lambda}^2\tau^3}{(1+\tilde{\lambda}\tau^2)^3} - \frac{4(1-\tilde{E})\tilde{\lambda}\tau}{(1+\tilde{\lambda}\tau^2)^2}$$
$$+ \frac{8(1-\tilde{\lambda})^2\tau}{(1+\tilde{\lambda}\tau^2)^2} - \frac{16(1-\tilde{\lambda})^2\tilde{\lambda}\tau^3}{(1+\tilde{\lambda}\tau^2)^3} + \frac{2(1-\tilde{\lambda})(1-\tilde{E})\tau}{(1+\tilde{\lambda}\tau^2)^2} \tag{22a}$$
$$+ \frac{(2-2\tilde{\lambda})\tau}{1+\tilde{\lambda}\tau^2}\left\{\frac{2(1-\tilde{\lambda})}{1+\tilde{\lambda}\tau^2} - \frac{4(1-\tilde{\lambda})\tilde{\lambda}\tau^2}{(1+\tilde{\lambda}\tau^2)^2} + \frac{1-\tilde{E}}{1+\tilde{\lambda}\tau^2} + \frac{4(1-\tilde{\lambda})\tau^2}{(1+\tilde{\lambda}\tau^2)^2}\right\},$$

and

$$s_2 = \frac{8(1-\tilde{E})\tilde{\lambda}^2\tau^2}{(1+\tilde{\lambda}\tau^2)^3} - \frac{2(1-\tilde{E})\tilde{\lambda}}{(1+\tilde{\lambda}\tau^2)^2} + \frac{2(1-\tilde{\lambda})(\tilde{E}-1)}{(1+\tilde{\lambda}\tau^2)^2} - \frac{8(1-\tilde{\lambda})(1-\tilde{E})\tilde{\lambda}\tau^2}{(1+\tilde{\lambda}\tau^2)^3}$$
$$+ \frac{1-\tilde{E}}{1+\tilde{\lambda}\tau^2}\left\{\frac{2(1-\tilde{\lambda})}{1+\tilde{\lambda}\tau^2} - \frac{4(1-\tilde{\lambda})\tilde{\lambda}\tau^2}{(1+\tilde{\lambda}\tau^2)^2} + \frac{1-\tilde{E}}{1+\tilde{\lambda}\tau^2} + \frac{4(1-\tilde{\lambda})^2\tau^2}{(1+\tilde{\lambda}\tau^2)^2}\right\}, \tag{22b}$$

.... etc.

Now, in following the Eq. (10), we have

$$\frac{s_0}{\lambda_0} = \frac{s_1}{\lambda_1} \longrightarrow \tilde{E}_0 = 1, \ -2\tilde{\lambda}+3,$$

$$\frac{s_1}{\lambda_1} = \frac{s_2}{\lambda_2} \longrightarrow \tilde{E}_1 = 1, \ -2\tilde{\lambda}+3, \ -6\tilde{\lambda}+5,$$

$$\frac{s_2}{\lambda_2} = \frac{s_3}{\lambda_3} \longrightarrow \tilde{E}_3 = 1, \ -2\tilde{\lambda}+3, \ -6\tilde{\lambda}+5, \ -12\tilde{\lambda}+7, \tag{23}$$

$$\frac{s_3}{\lambda_3} = \frac{s_4}{\lambda_4} \longrightarrow \tilde{E}_4 = 1, \ -2\tilde{\lambda}+3, \ -6\tilde{\lambda}+5, \ -12\tilde{\lambda}+7, \ -20\tilde{\lambda}+9,$$

...etc.

If we generalize the above expressions, we can find the bound state energy eigenvalues of the present system with a time-dependent mass

$$\tilde{E}_n = -n(n+1)\tilde{\lambda} + 2n + 1, \quad n = 0,1,2,\ldots. \quad (24)$$

After using Eq. (16), we can find the related energy spectrum in a more explicit form:

$$E_n = -\frac{n(n+1)}{2}\lambda + \frac{1}{2}(2n+1)\omega, \quad n = 0,1,2,\ldots. \quad (25)$$

Notice that the well-known usual case can be simply recovered when $\lambda \to 0$. Eq. (25) turns to become the usual energy spectrum for the constant mass oscillator.

In Figure 1, we plot the energy spectrum $E_n$ versus time dependent mass parameter $\lambda$ for different energy states $n = 0,1,2,3$ when the system oscillates with the frequency $\omega = 10$ Hz. In Figure 2, we also plot $E_n$ versus parameter $\lambda$ for $n = 1$ with different frequencies $\omega = 10Hz, 12Hz, 14Hz$. It is seen that the behavior is linear due to the presence of the correction term coming from the time-dependent mass. Figure 3 shows the behavior of the energy spectrum $E_n$ with the energy states $n$ for different oscillating frequency values $\omega = 10Hz$, $\omega = 20Hz$ and $\omega = 30Hz$ Finally, Figure 4, shows the energy $E_n$ with the frequency $\omega$ for the states $n=1,2$ and 3. The energy is linear with the oscillation of system.

Now, let us obtain the wave function of the Eq. (19). This equation is in the form of confluent hypergeometric differential equation:

$$x(x-1)y'' + [(a+b+1)x - c]y' + aby = 0$$

whose solution for $c$ not a positive integer, and $|x|<1$ that can be expressed as

$$y(x) = k_1 F(a,b;c,x) + k_2 x^{1-c} F(a-c+1, b-c+1; 2c; x)$$

where $F(a,b;c,x)$ is the hypergeometric function. After finding the solution of the equation and with second term in place, we can easily write the wave function of the Eq. (19) as follows:

$$\varphi_m\left(\tau, \frac{\lambda}{\omega}\right) = N_m \exp\left(-\frac{\lambda}{2\omega}\right)\left(1 + \frac{\lambda}{\omega}\tau^2\right) {}_2F_1\left(a,b;-2,1+\frac{\lambda}{\omega}\tau^2\right) \quad (26)$$

where $a = b = -\tilde{\lambda} - \frac{1}{2} \pm \sqrt{(2\tilde{\lambda}+1)^2 + 4(1-\tilde{E})}$, and $N_m$ are the normalization constants.

## 4. Discussion and Results

We studied the Feinberg-Horodecki equation with the time-dependent harmonic oscillator potential for a particle whose mass is time dependent. We used a general form for the mass in the oscillating form $m(t) = m_0/(1+\lambda t^2)$. The related stationary state energy eigenvalues and the corresponding wave functions are found for the present model by means of the time-like AIM. It is obvious that our results can be easily reduced to the solution of the FH equation for a constant mass case once we set $\lambda \to 0$. In fact, the analytical results obtained indicate that the time-dependent harmonic oscillator potential coherent states with time-dependent mass seem to be a key to understanding of the coherent formation of the specific growth patterns in the biological systems.

We also showed the influence of the mass parameter $\lambda$. on the energy spectrum $E_n$. In accordance to Eq.(25) and Figures 1 and 2, this behavior looks linear for different energy states. For the sake of completeness, in Figures 3 and 4, we have also presented the behavior of the energy with $n$ and $\omega$, respectively.

Finally, we want to remark that the energy eigenvalues in the present model are still stationary states and very like ones obtained using position-dependent mass.

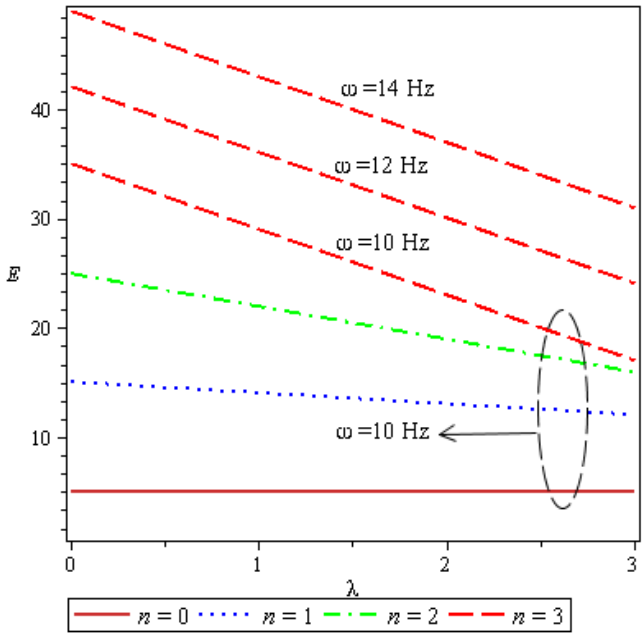

**Fig. 1** The stationary energy states $E_n$ is plotted against $\lambda$ when the frequency $\omega = 10\ Hz$

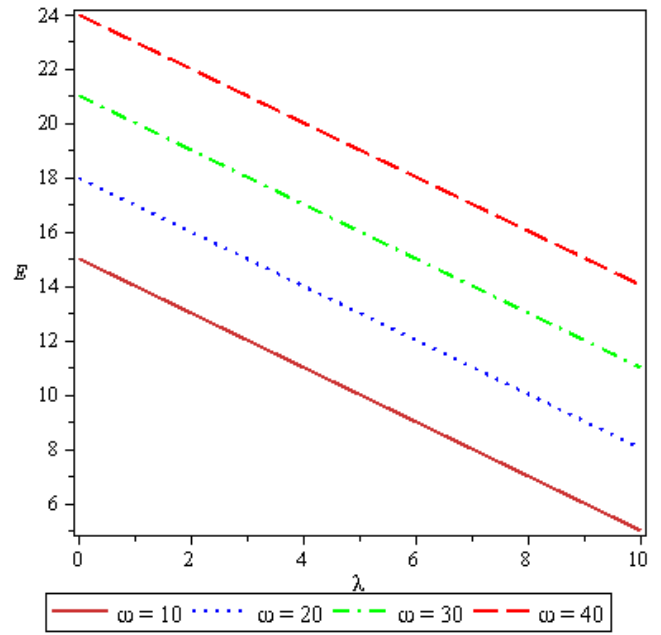

**Fig. 2** The excited energy state $E_1$ is plotted versus $\lambda$ for various frequency values $\omega = 10Hz$, $\omega = 20Hz$ and $\omega = 30Hz$

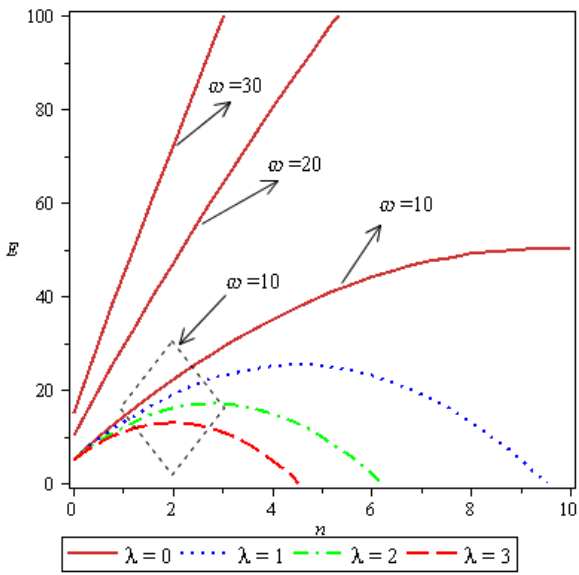

**Fig. 3** The energy states $E_n$ is plotted against $n$ for

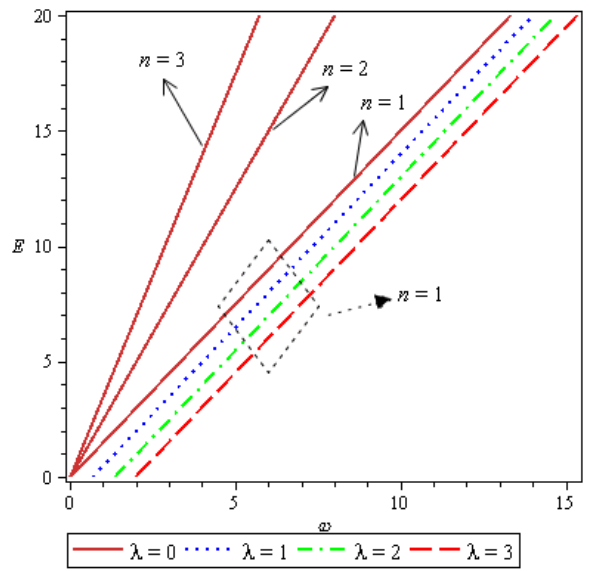

**Fig. 4** The energy states $E_n$ is plotted versus the

| various frequency values $\omega=10Hz$, $\omega=20Hz$ and $\omega=30Hz$ | frequency $\omega$ for various states $n=1,2$ and 3 |
|---|---|